\documentclass[conference]{IEEEtran}
\IEEEoverridecommandlockouts
\usepackage{cite}
\usepackage{amsmath,amssymb,amsfonts}
\usepackage{algpseudocode}
\usepackage{algorithm}
\usepackage{graphicx}
\usepackage{textcomp}
\usepackage{tabularx}
\usepackage{xcolor}
\usepackage{url}
\usepackage{csquotes}
\usepackage{listings, lstautogobble}

\def\BibTeX{{\rm B\kern-.05em{\sc i\kern-.025em b}\kern-.08em
    T\kern-.1667em\lower.7ex\hbox{E}\kern-.125emX}}

\begin{document}

\title{Grand Perspective: Load Shedding in Distributed CEP Applications
\thanks{This work was supported by the German Research Foundation (DFG) under the research grant "PRECEPT II" (BH 154/1-2 and RO 1086/19-2).
}
}

\author{\IEEEauthorblockN{Henriette Röger}
	\IEEEauthorblockA{\textit{University of Stuttgart} \\
		Stuttgart, Germany \\
		henriette.roeger@ipvs.uni-stuttgart.de}
	\and
	\IEEEauthorblockN{ Sukanya Bhowmik}
	\IEEEauthorblockA{\textit{University of Potsdam} \\
		Potsdam, Germany \\
		sukanya.bhowmik@uni-potsdam.de}
	\and
	\IEEEauthorblockN{Kurt Rothermel}
	\IEEEauthorblockA{\textit{University of Stuttgart} \\
		Stuttgart, Germany \\
		kurt.rothermel@ipvs.uni-stuttgart.de}}

\maketitle

\begin{abstract}
	In distributed Complex Event Processing (CEP) applications with high load but limited resources, bottleneck operators in the operator graph can significantly slow down processing of event streams, thus compelling the need to shed load.
	A high-quality load shedding strategy that resolves the bottleneck with high output quality evaluates each event's importance with regards to the application's final output and drops less important events from the event stream for the benefit of important ones. 
	So far, no solution has been proposed that is able to permit good load shedding in distributed, multi-operator CEP applications. On one hand, shedding strategies have been proposed for single-operator CEP applications that can measure an event's importance immediately at the bottleneck operator, only, and thereby ignore the effect of other streams in the application on an event's importance. On the other hand, shedding strategies have been proposed for applications with multiple operators from the area of stream processing that provide a fixed selectivity which is not given in the conditional CEP operators.
	We, therefore, propose a load-shedding solution for distributed CEP applications that maximizes the application's final output and ensures timely processing of important events by using a set of CEP-tailored selectivity functions and a linear program, which is an abstraction of the CEP application. 
	Moreover, our solution ensures a quality optimal shedder configuration even in the presence of dynamically changing conditions. With the help of extensive evaluations on both synthetic and real data, we show that our solution successfully resolves overload at bottleneck operators and at the same time maximizes the quality of the application's output.
\end{abstract}


\section{Introduction} \label{sec:introduction}
Complex event processing (CEP) uses continuous queries to detect patterns of interest in streams of primitive events.
\textit{Operators} implement these patterns and together form an operator graph. 
Under high load, 
an operator can quickly become a bottleneck, which leads to queuing of events and slows down the pattern detection. 
To resolve a bottleneck, parallelization\cite{mayer_predictable_2015, lohrmann_elastic_2015, roger2019combining, HenrietteBigData, russo_reinforcement_2019}, and re-placement \cite{cardellini_optimal_nodate, ottenwalder_migcep:_2013} require additional resources. 
Yet, if resources are limited, for example, in edge or fog processing, or given a limited budget for cloud resources, load shedding solutions \cite{slo2019espice, slo_pspice_2019, katsipoulakis2018concept} permit timely event processing while maximizing the output quality.
In load shedding, a load shedder drops events from an operator's input stream. Dropping those events shortens the input queue and reduces queuing latency.
Yet, dropping events risks to miss patterns. Missed patterns decrease an application's output quality. 
 However, non-critical applications permit moderate load shedding without compromising quality. 
 For example, a data center frequently sends CPU measurements. 
 If the pattern implements a load-increase detector  and the load shedder drops some, not all, of the CPU measurements, the pattern can still detect a load increase.
 
To minimize quality loss, a load shedder prioritizes important events where an important event is an event that has a high matching probability.
 Hence, an event's importance correlates with the event's impact on the \textit{application's output}. In CEP, an event's importance further depends on availability of the other events needed for a match. CEP operators are, therefore, \textit{conditional operators}. 

To determine an events importance, we need to distinguish single- and multi-operator CEP applications. 
Existing solutions apply a \textit{local perspective} \cite{slo2019espice, slo_pspice_2019, katsipoulakis2018concept, SloBR22, conf/debs/SloBR20, GeorgiaTechCollab, slo2023gspice}. The shedder rates an event's importance based on its local matching probability and thus the probability that the event contributes to an output event at the bottleneck operator.
This perspective is sufficient in single-operator CEP applications. Here, the operator's output is also the application's output as it is directly consumed by the sink.
In a CEP application with multiple operators, the application's output, i.e., the output the sinks consume, are the events emitted by the \textit{last} operators in the graph. 
The events that the bottleneck operator emits might be further processed by downstream operators before they are consumed by the sinks.  
To maximize the application's output, the local perspective 
is thus insufficient as quality indicator. 

The shedder rather has to consider which input events at the bottleneck operator will \textit{(1)} match at the bottleneck operator and \textit{(2)} how the resulting complex events will lead to further matches at those downstream operators.
Please note that in CEP operators, the ratio of input- to output events is non-linear. 
Most stream processing applications assume the number of output events to be proportional to the number of input events\cite{herodotou_survey_2020}. 
Yet, a CEP pattern requires multiple events to detect a match. Whether an event leads to a detected pattern depends on the availability of other events that are needed for a match.
We thus have a stateful selectivity \cite{zhao_load_2020} where the output rate depends not only on the event but on the history of detected events. 

Hence, a load shedder has to consider the downstream propagation of events rather than the bottleneck operator's immediate output to prioritize events for shedding. In doing so, the shedder also needs to consider availability of other events needed for those downstream matches. 
It thus needs a \textit{global perspective}.

\begin{figure}
\includegraphics[width=\linewidth]{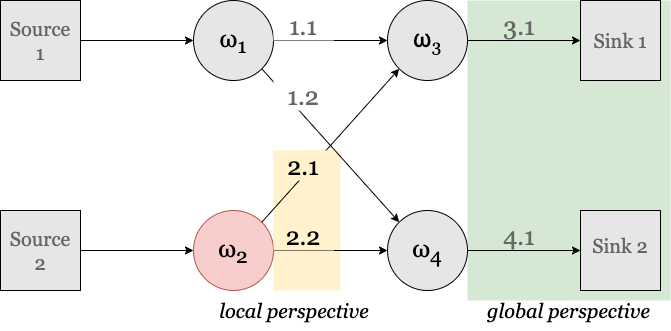}
\caption{An operator graph with two sources, two sinks and four operators. 
Operators $\omega_1$ and $\omega_2$ emit different outputs, denoted as 1.1 and 1.2 and 2.1 and 2.2 respectively. 
If $\omega_2$ becomes a bottleneck, a local shedder would maximize the outputs 2.1 and 2.2, while.a global shedder would maximize the outputs 3.1 and 4.1. 
To this end, it further needs knowledge about $\omega_1$'s outputs 1.1 and 1.2 and how operators $\omega_3$ and $\omega_4$ combine them with the bottleneck's outputs.
\label{fig:google_borg_application}}
\end{figure}

Figure \ref{fig:google_borg_application} depicts the two perspectives and gives a more concrete example that motivates the need for a global perspective.
The figure shows an operator graph with four operators, two sources and two sinks. Operators $\omega_1$ and $\omega_2$ emit events along two different edges. Along which edge an operator forwards an output depends on the pattern that has been matched. 

Assume $\omega_2$ to be a bottleneck. 
It emits events to Operators $\omega_3$ and $\omega_4$
where they each merge with events from $\omega_1$. 
To decide which input events to drop, the load shedder needs to know first, which input events match a pattern at $\omega_2$ at all and second, how likely is it for $\omega_2$'s output events to match at $\omega_3$ and $\omega_4$ respectively. 
A shedder with a local perspective maximizes the number of events $\omega_2$ emits, only.
Hence, it focuses equally on events sent along Edges 2.1 and 2.2. However, this ignores that Operators $\omega_3$ and $\omega_4$ additionally require outputs from $\omega_1$ and thus the availability of those events becomes important, too, to maximize the application's output.
Consider, e.g., availability of events on Edge 1.1 is low but on Edge 1.2 is high. Then the shedder at $\omega_2$ should prioritize events so that more are sent along 2.2 than 2.1. 
 Hence, it first needs to know which events will be sent along which edge, and second how many events are currently needed at operators $\omega_3$ and $\omega_4$, given the varied availability of events on Edges 1.1 and 1.2.
 The figure shows that the global perspective aims to maximize the number of outputs of operators $\omega_3$ and $\omega_4$ rather than $\omega_2$. This perspective requires knowledge about availability of events on Edges 1.1 and 1.2.
 In the worst case, assume no events along Edge 1.1 at all. In this case, the shedder should entirely focus on emitting events to Edge 2.2. A local perspective would still balance both, a global perspective knows about the lack of events on Edge 1.1 and prioritizes Edge 2.2, whilst shedding events that lead to events sent to Edge 2.1.

Finally, as CEP applications are usually long running and experience changes both in load and in the distribution of values in the event, a load shedder needs to frequently update what it sheds and how much to stay quality-optimal and latency compliant throughout the application's runtime.

To sum up, available solutions are insufficient for quality-optimal load shedding in distributed CEP applications with multiple operators. 
They either focus on local optimality \cite{slo2019espice} or consider a distributed application~\cite{tatbul_staying_2007} without conditional operators as given by CEP .
Yet, in the growing trends of edge and fog computing and IoT applications that are distributed yet resource limited, we see a strong need for a solution for quality-optimal load-shedding in geographically distributed CEP applications. 
The specific requirements for this solution are: 
\begin{enumerate}
	\item The solution  is applicable for conditional CEP operators.
	\item The solution defines the shedding so it maximizes the overall output of a distributed CEP application.
	\item The solution handles changes in the input and adapts to keep a latency bound.
\end{enumerate}

As a result, our contribution is a load-shedding solution for distributed CEP applications that  can maximize the application's output even under conditional operators distributed onto multiple nodes.  We, therefore, propose a set of selectivity functions tailored for CEP applications. We further propose a fine-grained shedder model and  a definition of a linear program to compute the optimal shedder configuration. The solution updates the load shedding if the application's load and value distribution changes at runtime. Finally, we evaluate our implementations on a real and synthetic data set.

\section{Background} \label{sec:background}
\subsection{Distributed Complex Event Processing}

A CEP application derives higher-level information from streams of primitive events. To this end, it applies \textit{continuous queries} that look for \textit{patterns} in the input streams. 

An event is a tuple \textit{(t, $\mathbb{A}$, ts)} with a type t, a set of attributes $\mathbb{A}$, and a timestamp ts. 

The attributes are the event's payload. 
\textit{Operators} implement the patterns of the queries and are assembled into a directed, acyclic graph that defines the flow of input- and output events from the events' \textit{sources} to the application's \textit{sink}.

Each operator defines three things: 
(i) the patterns it searches for in the incoming streams, (ii) the select-function that defines how to build complex events for pattern matches, and (iii) the input streams it consumes. 
The pattern is a combination of \textit{basic operators} that define the pattern's predicates.  
An operator verifies the pattern's predicates on incoming streams of events and emits an output if all predicates evaluate to true. 
If a pattern's predicates evaluate to true, a complex event is emitted, following the operator's selection function. 
We say an event \textit{contributes} to a complex event, if it belongs to the set of events that together fulfill the pattern's predicate.
Basic CEP operators to build patterns as enlisted in \cite{liu2011highperformance, chakravarthy1994snoop} are AND, SEQ, OR, negation thereof. 
We assume \textit{monotonic operators} as done by Zhao et al \cite{zhao_load_2020}: if a share of the input load is shed, the output is less than or equal to the output without shedding. 

The select-function defines how higher-order information is derived from matching events i.e., how a complex event is built. 
A complex event can be a derivative of the contributing events. Here, the complex event consists of the input events' attributes or a derivative thereof. More formally, $\mathbb{A}.e_{out} \subseteq \displaystyle\cup_{e_in} \mathbb{A}.e_{in}$ or  $\mathbb{A}.e_{out} = f(\displaystyle\cup_{e_in} \mathbb{A}.e_{in})$.
The output can also be a static event, e.g., a standard notification that is emitted once a match is detected or a composition of the two.

A CEP application can implement multiple queries that share resources and operator implementations.  The queries are split up into its individual patterns that are then distributed onto the operators. Each operator implements a share of the query's patterns and forwards its matches to subsequent operators.  
The operators can then be distributed onto multiple nodes which leads to a distributed CEP application.
The split of the queries into its patterns can be motivated by processing capacities of processing nodes, security requirements, reduction of network load, re-use of intermediate results for multiple queries.
How the patterns of a query are split up and assigned to operators is defined by the application's topology, placement, and query optimization decisions. Query-rewriting to optimize distributed CEP applications is well researched \cite{schultz2009distributed}.  
In this paper, we assume the split of the query's patterns onto operators and the operator placement to be given.

Formally, we denote the set of operators as $\Omega$, an individual operator as $\omega$.   We denote these patterns as $\gamma$, and the set of all patterns an operator implements is $\omega_\Gamma$.

If an operator implements multiple patterns, an event is applied on each pattern independently. As opposed to many stream processing applications (SP applications) that, e.g., aggregate values over a sliding window to compute averages or top-k lists, CEP operators search for patterns and emit an output as soon as the match is detected. 

\subsection{Load Shedding in Distributed CEP Applications} \label{subsec:shedder_model}

Shedding can be necessary if we observe operator overload. 
We  use load shedding to stick to a latency bound, not just to avoid back-pressure. 
The latency bound includes processing and queuing time, i.e., the time an event enters the node's queue until it is processed.

We target distributed CEP applications with a bottleneck operator. We assume each operator to have a latency bound. A bottleneck operator is an operator that risks bound-violations.
The shedder is placed at the bottleneck operator. %
In this paper, we focus on shedding events, not partial matches. Partial match dropping requires internal knowledge of the operator's state \cite{slo_pspice_2019}. However, in a distributed CEP application, internal state is not always available. 
Hence, we focus on shedding events.
A \textit{shedder configuration} defines which events to shed and how many.

\paragraph{Important Events and Quality Metric}
The key concept to maximize the quality is to understand the \textit{importance of events.}
The goal of a shedder is to ensure timely processing of those important events. To refine importance,  we distinguish three outcomes for an arriving event.
First, the event completes a partial match that is stored in the operator's state and a new complex event is emitted. Second, the event adds to a partial match and proceeds a state machine's state or opens a new one. Third, the event does not proceed any state machine nor fulfills an initial state to open a new one.
Clearly, the former two outcomes are more important as both lead or can lead to emission of a complex event. 
Informally, contributing events are important.
The challenge is to determine how likely it is for an event to contribute the moment the event arrives at the shedder.
 As a result, when a shedder in a CEP application defines an event's importance, the shedder has to consider an event's context at arrival and future events.

\section{Problem Statement} \label{sec:problem_statement}

Our problem statement is twofold. 
First, we define the problem of finding an \textit{optimal shedder configuration $c^*$} for a distributed CEP application.
Second, we add the notion of time, and define the problem of finding an optimal shedder configuration for changing \textit{workloads}.

\subsection{Optimal Shedder Configuration}
An operator's shedder configuration defines which input events to shed under which conditions. 
A \textit{feasible} shedder configuration is a shedder configuration $c_{f,t}$ at time $t$ that ensures events to be processed within a latency bound. 
The set of feasible configurations for an operator is defined as $\mathbb{C}_{f} \subset \mathbb{C}: \forall c \in \mathbb{C}: \text{c is  feasible}$.

The \textit{optimal shedding-configuration $c^*$} is a feasible configuration $c^*\in \mathbb{C}_f$ that further maximizes the application's recall. 
Please note that we assume monotonic queries as defined in Zhao \cite{zhao_load_2020}. 
Our goal is thus to maximize 
the application's output $\mathbb{O}_c$ under shedding.
The output  $\mathbb{O}$ is the sum of event arrival-rates $\lambda_s$ at the application's sinks. 
It is formally defined by $\mathbb{O} = \sum_{s \in Sinks}\lambda_s$. Hence, we want to maximize how many events arrive in average and per second at the sinks. 
Given the conditional properties of a distributed CEP application, computing $\mathbb{O}_c$  demands global knowledge. 

Our objective is to find  $c^*\in \mathbb{C}_f$ s.t. $\mathbb{O}_c$ is maximal.

\subsection {Changes in Workload}
In a long running CEP-application, the application's workload can change. The \textit{application's workload} is described by attributes of application's streams where a \textit{stream} $s_e$ along an edge $e$ of the operator graph is an unbounded set of events that flows from a source or an operator to another operator or a sink.
The attributes are $\lambda_{\omega}$, the overall rate of events per second arriving at the operator or sink, $\lambda_{T,\omega}$,  the rate of events per second of a specific event-type T in this stream, and the ratio $\lambda_{T}$/$\lambda_\omega$ for each type.
 If any of those attributes change, we speak about a change in the application's workload.  For example, one source emits more events of a certain type, changing the ratios. 
 Hence, an operator's workload can change, even if the overall arrival rate $\lambda_\omega$ is stable.
 Formally, we define the workload as described by the rates and ratios as \textit{stream characteristics} $SC$. Changes in the workload of the application are thus a change  of the application's stream characteristics.

 When we search for an optimal shedding configuration, we thrive for optimality given $SC_t$ at point in time $t$. If the $SC$ change, the configuration might become sub-optimal and we need to search for a new optimal configuration $c^*_{t+1}$. 
Further, feasibility is time dependent because of the dynamics in a CEP application. A configuration $c_{f,t}$ can be feasible at a point in time $t$, and due to changes in the workload, become infeasible at $t+1$. 

\subsection{Summary Problem Statement}
To sum up, our two goals are as follows: 
First, to find an optimal shedder configuration $c^*$ for a bottleneck-operator $\omega$, given a required latency bound $L_{max_\omega}$ and the operator's current input rate $\lambda_{\omega}$ and the application's stream characteristics $SC$.
Second, to react if a configuration becomes insufficient to meet the latency bound or becomes sub-optimal due to changes in the application's $SC$.

Our solution has to enable a global perspective to the shedder, permit fine-grained shedding for operator's with multiple patterns, and handle changes in the application's $SC$s. 
Our solution therefore first enables global-perspective shedding to maximize $\mathbb{O}$.
To this end, it uses a model that considers specifics of conditional operators as it uses CEP-tailored selectivity functions that we propose.
We then use this model to compute output-optimal shedder configurations.
Finally, we introduce a reaction-mechanism to update the configuration if $SC$ changes.

\section{Running Example} \label{subsec:running_example}
To explain our solution in detail, we extend the topology introduced in Section \ref{sec:introduction} to a running example. In particular, we introduce exemplary patterns that we later refer to in the discussion.

Figure \ref{fig:google_borg_application} shows an operator graph with four operators. 
We create a multi-pattern example application from this topology.
The application analyzes performance data from a google-borg cluster\footnote{\url{https://github.com/google/cluster-data}}.
We choose the 2019 data version.
The application has two sources  that emit CPU-measurements from different machines in a google-borg cluster.
We group the machines into four types by their $machine-id$. 
The goal of our simple example-application is to report outputs if CPU-measurements from both sources and different machine types are available.

The application implements this with two queries.
 Each CPU-measurement is an event E. Each query searches for six events E1 through E6 that fulfill the given conditions: 

\begin{lstlisting}[language=SQL] 
	Select E1, E2, E3, E4, E5, E6 
	Where E1, E2, E3 from Source 1
	And E4, E5, E6 from Source 2 
	And Seq(E1.type=0, E2.type=0, E3.type=1) 
	Within 10 sec 
	And Seq(E4.type=0, E5.type=0, E6.type=1) 
	Within 10 sec
\end{lstlisting}
\begin{lstlisting}[language=SQL]
	Select E1, E2, E3, E4, E5, E6 
	Where E1, E2, E3 from Source 1 
	And E4, E5, E6 from Source 2 
	And (E1.type=1, E2.type=2, E3.type=3) 
	Within 10 sec 
	And Seq(E4.type=1, E5.type=2, E6.type=3) 
	Within 10 sec
\end{lstlisting}

Both queries first require a pattern of events with certain types that stem from same source. When events  match this pattern from both sources, a match is forwarded to the queries' sink.
The first query requires a sequence where the events have to arrive in a specific order, written as Seq(E1.type; E2.type; E3.type).
The second query is less strict about the order and requires three types to appear within the given time limit.

We split up this query into patterns to have a distributed CEP application.
We choose to co-host the type-comparing sub-operators as they both work on the same property of the input, hence processing them together reduces communication and serialization effort. 
There is a broad range of literature available on how to split and place queries, e.g., \cite{hirzel_catalog_2014, cardellini_optimal_2017}. 
We exemplarily show the patterns implemented for $\omega_2$ and $\omega_4$. $\omega_1$ and $\omega_3$ are re-written from the original queries, respectively.

\begin{lstlisting}[language=SQL, escapeinside={(*}{*)}]

	Operator (*$\omega_2$*):
	Pattern 1: 
	Select E4, E5, E6 
	Where Seq(E4.type=0, E5.type=0, E6.type=1) 
	Within 10 sec. as CE.source = Q21
	Pattern 2: 
	Select E4, E5, E6 
	Where E4.type=1, E5.type=2, E6.type=3) 
	Within 10 sec as CE.source = Q22

	Operator (*$\omega_4$*): 
	Select CE1, CE2 as Output Sink 2 
	Where CE1.source =Q12, CE2.source=Q22
\end{lstlisting}
We group the output of $\omega_1$ and $\omega_2$ into a single, complex event CE and assign them a source attribute with source operator and pattern id, e.g. Q22, for $\omega_2$, pattern 2.
The $\omega_3$ and $\omega_4$ implement an AND pattern to match on these source attributes.

\section{Solution Overview} \label{sec:solution}
We search for the shedder configuration $c^*$ that keeps a latency bound ${B^*}$ at a bottleneck operator $\omega$ and maximizes the application's output $\mathbb{O}$.
 A configuration $c^*_t$ is optimal with regards to a $SC_t$. 
 We therefore further search for an update mechanism that provides  $c^*_{t+1}$ if $SC_{t+1}$ differs from $SC_t$ such that  $c^*_t$ is sub-optimal at $t+1$. 
We present our solution in four steps:

\begin{enumerate}
	\item We introduce fine-grained shedder model applicable for distributed, multi-query CEP applications.
	\item We propose an application model that uses CEP-applicable selectivity functions to evaluate the quality of shedding configurations. 
	\item We introduce the average processing-rate as metric and show why it is needed for shedding at high precision on multi-operator nodes.
	\item We transfer the problem statement into a linear program (LP) and show how to use it to find $c^*$
\end{enumerate}

Our solution solves one bottleneck operator at a time. We therefore explain it w.l.o.g. assuming  $\omega_2$ to be the bottleneck.

\section{ System Model} \label{sec:system_model}

\subsection{Event Types} \label{subsec:bg_streams_and_classes}
A shedder distinguishes important and unimportant events. To this end, it needs to categorize input events accordingly. 
Our load shedder sheds events based on their \textit{type}. 
In the running example, we define the event's type based on the event's machine id. 

For now, we assume that we are provided with all types. 
However, it is an interesting future work to obtain these types from the query. 
An edge in the operator graph can transport streams of events from multiple event types. 
Operators apply the patterns $\gamma$ on their input streams and generate output streams, which, given the notion of types, transfers to a mapping of input types to output types.

\subsection{Local vs. Global Load Shedder}
Compared to local solutions, in a distributed setting, a shedding decision requires information from multiple nodes. Slo et al. \cite{slo2019espice} propose a shedding table where the overloaded operator automatically chooses the correct amount to shed. Thus it is a local decision. 
However, in a distributed setting, we need to consider output rates at other nodes, too, to maximize the output under shedding. 
Thus, in a distributed setting, an operator $\omega$ cannot locally decide what and how to shed based on a plan. 
This would lead to sub-optimal solutions. 
Hence, we need to introduce an information-emission strategy and choose a centralized component that collects all data and computes for global optimality. However, this of course leads to slightly slower reaction time than a purely local solution where the operator only looks up the required shedding ratio for a specific input rate.

 If an operator emits a single type, only, the shedder only needs to maximize this output rate. This does not require global knowledge.
 We need global optimization if an operator emits multiple output types. 
Which type to shed considers the type's contribution to the application's \textit{final} output. The contribution to the operator's immediate output is insufficient. As seen in the running example, which input types to shed at $\omega_2$ depends also on the availability of outputs from $\omega_1$ as they will eventually be merged in $\omega_3$ and $\omega_4$. Hence, a more global knowledge about all operators involved is required.   

\subsection{Quality Metric}
We judge the quality of a shedder using the application's recall. 
The recall is the ratio of events the application did detect under shedding compared to the number of events that would be detected without shedding. We aim at maximizing the recall. 

\subsection{Query Granularity}
In this paper, we assume that all patterns are written at a type-granularity. For example, $\omega_1$ has the two patterns SEQ(0;0;1) and AND(1,2,3). We thus assume that the involved types are 0,1,2, and 3 and that a shedder can tell to which type an event belongs.

\subsection{Nodes, Operators, and Patterns}
 A \textit{node} is a physical machine and can host one or multiple operators. 
The placement problem of how to assign operators to nodes is well researched and seen as out of scope of this paper. 
We do not make any requirements on the node's hardware, e.g., homogeneity as it is common in many parallelization-solutions. 

Each \textit{operator} can further implement multiple \textit{patterns}, e.g., when an application implements multiple queries that use the same events. 
This set of patterns $\Gamma$  on an operator
is applied \textit{sequentially} on the incoming streams.

\subsection{A Fine-grained Shedder Model for CEP operators} 
\subsubsection{Probabilistic Shedder configuration}
It defines which \textit{share} of an input type to process and which share to drop. 
At an event's arrival,  the event is processed with the given probability and discarded else. 
We shed shares rather than all events of a  type (c.f. \cite{slo2019espice}) to comply with the conditional operators.  Consider the sequence  (0;0;1) as used in $\omega_1$ and $\omega_2$ in the example, and $\lambda_1: 1000e/s, \lambda_0:500 e/s$. Hence, we can shed a share of events of type 1 as we have fewer events of type 0 to match. However, we should not shed all events of type 1 as no more matches will be found.

\subsubsection{Pattern-wise shedder configuration}

An operator can implement multiple patterns. 
For each pattern, the importance of an event can differ for two reasons:
 First, the event's type is not used in a pattern. For  example, an event of type \enquote{2} is not used in the pattern 1 but in pattern 2. It is therefore unimportant for one but important for another pattern. 
Second, an event's type can be needed in both patterns, for example type \enquote{1} but the required amount differs.
This amount depends on the availability of the other event types needed for that pattern.
This can be caused by other required types being limited either at the same operator (e.g., few events of type \enquote{0} to match with type \enquote{1}) or by other events required for matches downstream, e.g., outputs from $\omega_1$ needed for a match at $\omega_3$ or $\omega_4$.. 
As a result, our shedder model has to be able to shed at different ratios per pattern.

Together with the probabilistic shedding motivated before, we therefore propose a shedder model where the shedder configuration assigns \textit{shares} to shed {per event type and  pattern}.
Formally, a shedder configuration $c_{\omega}$ is a set of assignments \{($\gamma, T) \rightarrow r_{\gamma, T}: r_{\gamma, T}\in [0,1] \forall \gamma \in \Gamma_\omega, \forall T \in \mathbb{T}_{in_{\omega}}$\} with $r_{\gamma,T}$ being the share to shed of all events of type $T$ at pattern $\gamma$ and $\mathbb{T}$ being the set of all types used in patterns on $\omega$. 
The shedder uses the configuration to shed each event of type $T$ at pattern $\gamma$ with the probability c($\gamma, T$).

\section{Finding $c^*$ }
Given the formal definition of a shedder configuration for an operator $\omega$, we now need to find configurations that ensure the required latency-bound yet maximize the application's (not just the operator's) output. We denote configurations that meet this criteria as $c^*$. 
Formally,  $c^*$  fulfills two properties: 
It is  \textit{feasible} and it maximizes the expected output $\mathbb{O}_c$. To verify these two properties of a configuration $c$, we need to be able to test feasibility and a $c$'s expected output.
In the following, we first describe how we compute whether a configuration is feasible using the concept of the \textit{average processing time}. We then describe how to compute a configuration's expected output using selectivity functions tailored for CEP operators. 

\subsection{Computing Feasibility} 
As described in Section \ref{subsec:shedder_model}, we assume a latency bound to be given for the bottleneck operator. 
The latency bound includes queuing latency $B_{waiting}$ and the actual processing time.  
Our shedder does not reduce an event's \textit{individual}  processing time.
We focus on controlling the expected average time $B$ an event spends at an operator. 
To achieve a latency bound with shedding, we focus on reducing the average processing time $p$. Controlling $p$ controls the queuing latency as a fast $p$ ensures fast processing of the queue. Yet, it also permits us to process events at those patterns where they are important and shed them where they are less important, following our fine-grained shedder model. In the following, we describe this concept in detail.

\subsubsection{Reducing the Average Processing Time}

We introduce the intuition behind the control of $p$ using a single event that arrives at the operator. 
We then apply queuing theory to extend our discussion to a general perspective that comprises all events arriving at the operator. 

 Intuitively, an event $e$ needs to queue $Q_t * p$ seconds. $Q_t$ is the number of events in the queue when $e$ arrives at the queue at time $t$  and $p$ is the operator's processing time.
 $p$ includes the processing time of all patterns implemented on the operator. 
 This processing time can differ per event. 
 For example, an event that immediately leads to a match has a shorter processing time than one that requires all internal state machines to be checked and matches at the last one. 
 Hence $p$ is the \textit{average}, also defined as $1/\mu$ with $\mu$ defined as the operator's average processing rate. 
 More formally, under the assumption of an M/M/1 queue model, we apply Little's Law for M/M/1 queues to compute the average time $B$ an event spends at the operator (Eq. \ref{eq:operator_delay}). If $B$ is now bounded as $B^*$, Equation \ref{eq:delay_delta}  shows how shedder configuration that leads to $p*$ is a feasible configuration given an arrival rate $\lambda$.  

\begin{equation}
\label{eq:operator_delay}
	B = \frac{1}{\mu-\lambda} = \frac{1}{\frac{1}{p}-\lambda}
\end{equation}

\begin{equation}
\label{eq:delay_delta}
	B^* = \frac{1}{\frac{1}{p*}-\lambda}
\end{equation}

Note that those equations are not defined for overloaded operators, i.e., where $\lambda \geq \mu$. 
Overloaded operators face infinite queuing. 

$\lambda$ is the average arrival rate at the operator. $\mu$ is the average processing rate. 
If for a given $\lambda$ and $\mu$, $B$ exceeds the operator's bound, we need to reduce it. 

To reduce $B$, we can either reduce $\lambda$ or increase $\mu$. 
Reducing $\lambda$ with shedding means dropping events at $\omega$'s input queue., i.e., not to process those dropped events at all.
However, as described, our model permits to process events at a subset of $\omega_\Gamma$, using pattern-wise shedding. 
Thus, we aim to increase $\mu$, the rate of events processed per second by reducing the average processing time  $p$.

When reducing $p$, we reduce the processing time \textit{averaged} over all events and patterns. $\lambda_{T}$ is the arrival rate of a type $T$, $ptime_\gamma$ is the average processing time of events at pattern $\gamma$.
We process a share of the incoming tuples of a type at a subset of $\gamma \in \Gamma$, denoted with $r_{T,\gamma}$. 
Equation \ref{eq:avg_p_time} shows how to compute $p$ under shedding.

\begin{equation}
	\label{eq:avg_p_time}
	p= \sum_{T \in \mathbb{T}}\frac{\lambda_{T}}{\lambda}*\sum_{\gamma \in \omega_\Gamma}x_{\gamma,T}*ptime_\gamma + (1-x_{\gamma, T})*0
\end{equation}

Shed events account to a processing time of 0,
processed events account for the pattern's processing time.

Now that we understand what parameter to influence and how to influence it, we can then compute $p*$ as shown in Equation \ref{eq:delay_delta}. In an overloaded operator, $p*$ must be sufficiently low to ensure $\lambda < (1/p*)$.

There can be multiple feasible shedding configurations. Hence, we further want to find those that maximize $\mathbb{O}_c$. 
To this end,  we need a model that lets us compute how a shedder configuration influences  $\mathbb{O}_c$. 

\subsection{An Application Model for CEP Applications}
To compute $\mathbb{O}_c$, we need to model the shedder's influence on $\omega$'s output and how this propagates through the operator graph until the sinks.
We aim to find the shedding configuration that maximizes the expected recall (i.e. maximize $\mathbb{O}$).

We propose an \textit{application model }that  is an abstraction over the operator graph and provides a connected set of selectivity functions.  
 The selectivity functions map the input arrival rates  $\lambda_{\omega,T}$ to the operator's expected output rates.  By connecting them according to the operator-graph structure, they can be used to compute $\mathbb{O}$. 
 
 Commonly, the selectivity function is a single-input, linear function that applies a factor $\lambda_{\omega,T}$ to compute $\omega$'s output.
 For example, consider $\omega_2$. It receives events of types 1,2, and 3 and emits events of types Q21 and Q22. 
 Hence, a linear selectivity function for input type 1 and output type Q21 could be $sel_{T 1, Q21} : \lambda_{Q21} = 0.5* \lambda_{T1}$, which specifies that every event of type 1 leads statistically to 0.5 output events. 
 While this is a good approximation for common stream processing functions, e.g., filter, or window-aggregations, where all elements in a window are aggregated to a single output, for CEP, it falls too short. 

Linear, single input selectivity functions are limited for the conditional operators of a CEP application. 
In conditional operators, an event's selectivity depends on the availability of other events. It can only lead to an output if all events required for the pattern match are available. The input to output ratio needs to consider the availability of all required events. 
 Hence, we face \textit{conditional selectivities}. For $\omega_2$ pattern 1, we need two events of type  0 and one of type 1 to emit an output. 
 Hence, the selectivity of an event of type 1  depends on the available events of type 0, e.g., $sel(T_1, Q21|T_0) = 0.7$. If no events of type 0  are available, no outputs will be produced: $sel(T_1, Q21|\bar{T_0}) = 0.0$. We thus search for selectivity functions that capture these dependencies. 
Formally, we write $\mathbf{{\lambda_{out}}} = sel(\mathbf{\lambda_{in}})$, saying that an operator's output rate is the function value of the vector of arrival rates.

We propose selectivity functions that follow the conditional nature of CEP queries.
 
\subsection{Selectivity Functions for CEP Operators}
Table~\ref{tab:buildModel} gives an overview of how we define selectivity functions for each basic operator. 
\begin{table*}[h]
\caption{Selectivity functions for CEP-query components. The capital letters are the \textit{arrival rates} of the respective input types.}\label{tab:buildModel}
	\begin{tabularx}{\textwidth}{|l|l|X|}
\hline
\textbf{Basic Operator} &\textbf{ Example} &\textbf{Model Representation (Selectivity Function) }\\ \hline
SEQ & A;B;A & min(A/2, B) \\ \hline
AND & A and B & min(A,B) \\ \hline
OR & A or B & max(A,B) \\ \hline
Property (closed set) & A and B with A.stock = IBM and B.stock = Apple & min(A\_IBM, B\_Apple) \\ \hline
\end{tabularx}
\end{table*}
Note that for those selectivity functions, we assume a consume-once consumption policy, i.e., each event can match exactly once per pattern.

To define selectivity functions for CEP operators, we exploit the fact that a CEP pattern consists of basic operators as introduced in Section \ref{sec:background}.
Those basic operators are: 
\begin{itemize}  
\item Sequence (A;B) with or without intermediate values
\item AND (A and B)
\item OR (A or B)
\end{itemize}
A pattern is an arbitrarily complex combination of those building blocks. 
For each of these basic operators, we propose a multivariable selectivity function and re-write the query into a connected graph of selectivity functions. To this end, we nest the selectivity functions the same way the basic operators are nested within the query. 

For example, the output of $\omega_2$ is consumed by $\omega_3$. $\omega_3$ further consumes the output of pattern 1 at $\omega_1$.
Equations \ref{eq:sel_omega1} to \ref{eq:sel_omega3} depict how we nest the respective selectivity functions.
If we replace $\lambda_{T_{Q11}}$and $\lambda_{T_{Q21}}$ using Equation \ref{eq:sel_omega1} and \ref{eq:sel_omega2}, we are able to predict the output of $\omega_3$ from the application's input consumed at $\omega_1$ and $\omega_2$ respectively. 

\begin{equation}\label{eq:sel_omega1}
	\lambda_{T_{Q11}} =min(\lambda_{T_0,\omega_1 }/2, \lambda_{T_1, \omega_1})
\end{equation}
\begin{equation}\label{eq:sel_omega2}
	\lambda_{T_{Q21}} =min(\lambda_{T_0,\omega_2 }/2, \lambda_{T_1,\omega_2})
\end{equation}
\begin{equation}\label{eq:sel_omega3}
	\lambda_{Sink 1} =min(\lambda_{T_{Q11}},\lambda_{T_{Q21}})
\end{equation}

If we now modify an arrival rate $\lambda_{\gamma, T}$ by applying a shedding ratio, we're able to tell the effect at the sink.

 So if we modify any of those arrivals with shedding (note that shedding in front of a pattern actually influences the pattern's arrival rates, yet we do not influence the operator's arrival rate in the first place), using equations similar to Equations
\ref{eq:sel_omega2} and \ref{eq:sel_omega3}, we can predict changes at the sinks.

\paragraph*{A remark on sequences}
We assume that in a distributed setting, events do not arrive in a global order, e.g., due to channel delays. Hence, we treat sequences like \texttt{AND}-operators in the selectivity function domain.
It is, however, possible, to enforce an ordering in the concrete operator implementation. This might require synchronized event timestamps and is out of scope of this work.

\subsubsection{Summary of the Application Model}
Our model provides for each pattern $\gamma$ a function $\mathbf{\lambda_{T_{in}}} \rightarrow \lambda_{T_{out}}$ that computes the expected average output rate  for a given vector of input-rates per type. 
Composing those patterns according to the operator-graph's topology provides a model to compute expected outputs - both in the middle of the graph and at its sinks. 
Hence, if a shedding configuration modifies a pattern's input rates $\lambda_\gamma$,  we can approximate $\mathbb{O}_c$ and find the configuration that maximizes it.

Now that we understand the properties of $c^*$ and how we use it to control the operator's latency bound, in the following we show how we compute $c^*$ using a linear program, and finally how we update $c^*$ if we observe changes in the application's load.

\subsection{Compute $c^*$ With a Linear Program}

In the following, we describe how we re-write the problem statement as a linear program (LP). We use the selectivity functions of the application model and the concept of average processing time $p$. 
Re-writing the problem statement lets us compute $c^*$ using a standard solver.
In the following, we explain the parts of the LP.
We show how we model the decision variables, the objective function, and the constraints.

Table \ref{tab:LP_params} gives an overview of the symbols we used. 

\begin{table}[t]
\caption{Symbols used in the LP}\label{tab:LP_params}
	\begin{tabularx}{\linewidth}{|l|X|}
\hline
\textbf{Symbol} &\textbf{Description} \\ \hline
$\omega, \Omega$ & an operator, the set of all operators of the application \\ \hline
$\omega_{\gamma}, \omega_\Gamma$ & a pattern $\gamma$ at operator $\omega$, the set of all patterns at operator $\omega$ \\ \hline
$\omega_{in}, \omega_{out}$ & the set of input and output types of $\omega$  \\ \hline
$\lambda, \mu, \rho$ & arrival rate, processing rate, utilization. Can be indexed with $\omega$ or $\gamma$, representing the respective values for the complete operator or for a single pattern.  \\ \hline
$x_{\gamma, T}, y_{\omega,\gamma}$ & The decision value (dv) in the LP that represents the drop-share and the computed output of the pattern $\gamma$ at operator $omega$. \\ \hline
$r_{\gamma,T}$ & The processing ratio at pattern $\gamma$ for an input type $T$. The share of inputs from T that will be processed at $\gamma$. (1-r) is the share of inputs to be shed.
 \\ \hline
 $ succ_\gamma$, $succ^*_\gamma$ & Successor-definition:is the set of those operators that directly consume output's of the the pattern $\gamma$. $succ^*_\gamma$is the set of transitive successors, i.e., the operator's successors and their transitive successors.    \\ \hline
\end{tabularx}
\end{table}

\subsubsection{Decision variables}
The LP has two types of decision variables (dvs). First, those variables that represent a shedding-ratio, which we denote with $x$, and second, those variables that represent a pattern's output, which we denote with $y$. 

We use the $x$-dvs for the shedder configurations and the $y$-dvs to maximize $\mathbb{O}$. 
 The solver assigns a value between 0 and 1 for each $x_{\gamma, T}$. It defines the share of the input type T to be shed for $\gamma$. 

\begin{figure}
	\includegraphics [width=\linewidth] {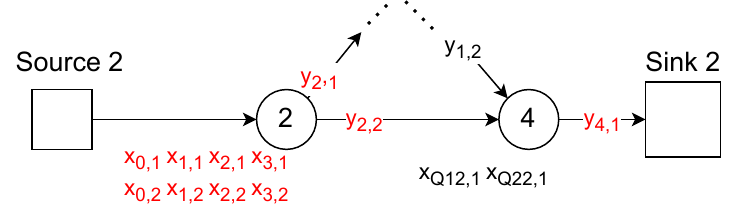}
	\caption{An excerpt from the example application. The figure depicts where which type of dv is relevant. The x-dvs denote the shedding ratio at each operator's input, the y-dvs denote each pattern's output. 
	If $\omega_2$ is the bottleneck operator, only the red highlighted dvs are to be found. The black dvs are fixed values, making the LP linear.}
	\label{fig:dv_figure}
\end{figure}

Figure \ref{fig:dv_figure} shall give an intuition how the dvs are placed in the operator graph. 
The figure shows an excerpt of the example-application's operator graph. 	The x-dvs are placed in front of the operators to depict the shedding ratios. The y-dvs are placed at the output-arrows to show that they are each pattern's output. If we see $\omega_2$ as bottleneck operator, the red dvs are those to be found by the LP. 

Equations \ref{eq:x_dvs}, \ref{eq:y_dvs} formally define the dvs. 
\begin{equation}\label{eq:x_dvs}
	x_{T,\gamma} \in [0,1] ~ \forall \gamma \in \omega_\Gamma, ~T \in \omega_{in}, ~\forall \omega \in \Omega
\end{equation}

\begin{equation}\label{eq:y_dvs}
	y_{\omega,\gamma} \in \mathbb{R}^+~ \forall \gamma \in \omega_\Gamma, ~\forall \omega \in \Omega
\end{equation}

The equations show that the number of $x$ variables is the product of the number of operators, the patterns per operator, and the number of input-types. 
The number of $y$ variables depends on the number of operators and the patterns per operator.

\subsubsection{Objective}
The LP's objective uses  those y-dvs that represent the outputs of the last operators before the sinks, i.e., we maximize the sinks' input.  
Every $y_{\omega,\gamma}$ represents the output rate of $\gamma$ at operator $\omega$ and becomes the input rate of the successor-operators. 
Of particular interest are thus those $y_{\omega,\gamma}$-dvs that become the sinks' input. These are the rates we aim to maximize to maximize $\mathbb{O}_c$. 

\begin{equation} \label{eq:global_objective}
	max \sum y_{\gamma}*weight_{sink} ~\forall \gamma :  succ(\gamma) = {sink}, \forall ~sink \in \mathbb{S}
\end{equation}

Equation \ref{eq:global_objective} gives the optimization objective of the LP. The total arrivals at all sinks shall be maximized. The total can be weighted to prioritize specific sinks.

\subsubsection{Constraints}
The LP has three types of constraints: (1) constraints that implement each pattern's selectivity function,  
(2) constraints that limit each operator's output to the operator's maximum processing rate, ensuring that an operator can never emit more events than it can process, even if more events arrive, and (3) a constraint that enforces the average processing time limit and thus implements the latency-requirement compliance.

\paragraph{Selectivity-function Constraints}
An operator's selectivity functions define the expected output rate for a given multi-dimensional input rate.
We introduce constraints into the LP that define this relationship. 
If the input rates are then reduced with shedding, the solver reduces the respective output rates, too. 

Equation \ref{eq:pattern_computation} shows how the y-decision variables $y_{\omega,\gamma}$ become the function value of the pattern's selectivity function. The input are the current arrival rates $\lambda_{in}$, reduced by the shedding ratio represented by the x-decision variable $x_{T{in},\gamma}$. 
 
 It is possible that multiple patterns emit the same output type which is reflected in Equation
 \ref{eq:output_computation}.
 

\begin{equation} \label{eq:pattern_computation}
	y_{\omega,\gamma} = sel_\gamma(\mathbf{\lambda_{T,\omega}} * (1-x_{ T,\gamma}))
\end{equation}

\begin{equation}\label{eq:output_computation}
	\lambda_{T,\omega} = \sum_{\gamma': \gamma' \in pred(\omega)}y_{\omega'\gamma'} 
\end{equation}

To use Equation \ref{eq:pattern_computation} for the LP, we need to re-write the selectivity functions accordingly. 
Table \ref{tab:buildModel} shows that this requires re-writing $min$ and $max$ statements.

\begin{equation} \label{eq:AND_example_1}
	y_{AND} \leq x_{AND,A}*\lambda _A
	\end{equation} 
\begin{equation}\label{eq:AND_example_2}
		y_{AND} \leq x_{AND,B}*\lambda _B
\end{equation}

 As an example, Equations~\ref{eq:AND_example_1} to \ref{eq:AND_example_2} show how we re-write an AND-operator with the selectivity function min(A,B) and the shedding ratios as conditions for the LP. 
 The output variable $y_{AND}$ needs to be equal or smaller than any input after shedding. 

\paragraph{Output constraints}
The second constraint says that an operator cannot emit faster than it can process. 
This limit is relevant for overloaded operators and further shows how shedding at an overloaded operator can increase the output rate: shedding increases operator's $\mu_{\omega}$.Hence, the upper bound for the output $y_{AND}$ rises, enabling higher output rates. 
\begin{equation}\label{eq:AND_example_3}
	y_{AND} \leq f*\mu_{\omega}
\end{equation}
Equation \ref{eq:AND_example_3} extends the example for its output constraint. 
The parameter \textit{f} takes into account that in case of a match, a pattern can also emit multiple outputs (of the same type). Usually, f is one. However, we introduce f to show how to cover cases where a match can lead to multiple outputs, similar to, e.g., a sentence-split operator that emits multiple words for each input sentence. 

\paragraph{Average processing-time constraint}
The third type of constraint ensures that $\omega$ sticks to $p*$. 
The Equation \ref{eq:avg_p_time} translates into an LP. 
For each LP-run, the arrival rates $\lambda$ and the processing time per pattern $ptime_\gamma$ are measurements taken from a system snapshot.

 
 \paragraph{A Short Remark on Linearity}
Equation \ref{eq:output_computation} shows that the arrival rates $\lambda_{T,\omega}$ are a function of the output rates represented by the y-dvs. 
Equation \ref{eq:pattern_computation} further shows how arrival rates are used to compute y-dvs. 
In Figure \ref{fig:dv_figure} we show how the output of $\omega_2$, $y_{2,2}$ becomes together with $y_{1,2}$, the input of $\omega_4$ as described in Equation \ref{eq:output_computation}.
If we now applied Equation \ref{eq:pattern_computation} for $\omega_4$, we multiplied dvs ($y_{1,2}, y_{2,2}$ with $x_{Q21,1}$ or $ x_{Q22,1}$), violating the linearity requirement of an LP.
However, the LP is solved for one bottleneck-operator at a time, so that all x-dvs for other operators are by default set to 0: $\forall x_{T,\gamma}, \forall \omega \in \Omega \setminus \omega^*: \gamma \in \omega_\Gamma \wedge : x = 0$ with $\omega^*$ being the bottleneck operator.
This ensures linearity in all conditions. It further keeps the amount of dvs per LP-run small. Independent of the graph size, the number of variables to assign a value is limited to those dvs affected by the current bottleneck operator. 
The number of x-variables to assign per LP-solving round is limited by $|T_{in}| * |\gamma|$ for the bottleneck operator. Currently, we assume for each pattern to emit a single output type, i.e., $|T_{in}|$ is upper bounded by the number of upstream predecessor-patterns or the number of types emitted by a source. 
For the y-variables, the following statement holds: $\forall y_\omega,\gamma: \gamma \in \Gamma_\omega \wedge\omega \notin succ*(\omega^*) : y = \lambda_{out_\gamma}$. 
This says that all output variables unaffected by shedding, as their operator is not a successor of $\omega^*$, get the fixed output value currently measured in the system.

We implement one LP for the complete application (which would be non-linear because the output of one operator $y$ becomes the input of the next, multiplied by $x$, the ratio at this operator) but make it a linear problem for each individual operator to solve.

 \section{Controlling $c^*$ and processing times at runtime}\label{sec:update}
 
  Now that we introduced the model to find $c^*$ for a given system snapshot, we use it to control the application's latency at runtime under changing stream characteristics.

We propose a three-step solution. 
First, we continuously monitor the application's stream characteristics and processing times.
Second, we verify whether significant changes happened that require a (re-) computation of $c^*$. 
Third, if a significant change is detected, we apply the LP for the latest monitored data to re-compute $c^*$ and forward the new configuration to the bottleneck operator. 

This continuous update allows our system to quickly react to changes in the stream characteristics or processing times and is a novelty as, e.g. \cite{slo2019espice} learn a model once, apply it to the application that then independently controls the queuing delay. However, it assumes a fixed utility for each type which cannot be assumed in a global setting where  we have inter-operator dependencies.

\subsubsection{Monitoring}

The CEP application measures the workload, the emission rates, and the operator's processing times at each operator. 
All measurements are made on a per-type level and on a sub-operator level to have sufficient fine-grained data.
Each operator computes a running average over an adjustable number of events. 
If this average deviates significantly from the last stored average,  it is forwarded to the shedding controller.
With this runtime monitoring, we omit initial profiling.

We start a (re-)computation of the optimal shedder-configuration in two cases:
First, if the newly reported average processing time exceeds $L_{max}$. 
Second, if the average processing time was below $L_{max}$ \textbf{and} shedding was active. 
The first case implies that more shedding is required. 
The second case, however, implies that the currently active shedding configuration sheds too much, reducing the quality unnecessarily. This can happen e.g., if the arrival rates shrink or the stream characteristic change since the last shedder configuration. 
In this case, a new configuration might include lower shedding ratios.
Both cases trigger the LP-solver.

As input rates, we use the upstream operator's output rates, not the measured arrivals of an operator. This is important to have the real workload, even if an operator is overloaded. An overloaded operator cannot count more events per second than it can process, yet, it is possible to have  $\lambda > \mu$ which we detect by summing up the upstream output.

\section{Implementation and Deployment}
\subsection{Metric Aggregation and Overhead Analysis}
One important question with regards to high-loaded systems is the additional load of the shedding solution. 
In geo-distributed CEP applications, low load on networks is preferable. This holds especially for bandwidth-limited connections, e.g., when using mobile devices with cellular connection. 
Hence, from a network-load perspective it seems preferable to compute the average processing time and the total emission count on the node and send only this aggregated value to the central component. 
Sending aggregated values can further be done event-based, i.e., only if significant changes are observed, or frequency-based, e.g., every five second. 
We implement event-based monitoring, where a node forwards an update whenever a deviation-threshold from an expected or formerly measured value is met. This can be very efficient given stable processing and emission rates. This, again, is given by stable stream characteristics, i.e., a low-variance distribution of event-types in the stream and a low variance in each type's arrival rate. However, if any of these variances increases, we face a higher update rate. The threshold defines the sensitivity to these variances. However, we still cannot tell in advance the expected load on the network. Further, to compare against a threshold, the node has to do the necessary aggregations. This can become a bottleneck on an overloaded node.
\subsection{Parameters for the LP}
The LP uses $\lambda$, $ptime$, and $\mu$ as parameters. For a single solver-run, those values are fixed, i.e., they are not decision variables. 
To obtain them, we take a snapshot from the running system and provide them as parameters to the solver. Hence, they are fixed per run, but adapt at runtime to changes in the stream characteristics.

\subsection{Managing internal state}
We implement the pattern matching with state machines. An open state machine is discarded if the pattern's time or count constraint is met and the final state is not achieved.
When a new event arrives at a node, the node verifies whether the event proceeds any of the open state machines. 
If this is the case, the event is stored and the machine proceeds to the next state. 
If the final state is reached, a complex event is emitted, applying the select-function on the state machine's stored events and the state machine is closed. 
If no matching state machine was found, the node verifies whether the event fulfills the initial condition to open a new state machine. 
We assume a consume-once per rule policy. 

\section{Evaluations}
In our evaluations, we select a single operator in a multi-operator CEP application as bottleneck operator. We assess how the global output-quality and the operator's processing time perform when shedding the input events using the LP-solution. 
As quality metric, we count an experiment's output, i.e., the sum of events arriving at both sinks. As latency metric, we measure if the selected operator sticks to the assigned processing-time limit with its shedding actions.

We evaluate our solution in terms of quality, latency compliance and scalability. For quality, we measure differences of a local and our global perspective for different data sets with regards to the application's output quality. For latency compliance, we measure whether the operator, throughout runtime and in total, maintains the required processing time. For scalability we scale up the query and test how properties of the query influence the solution's runtime.

\subsection{Local and Global}
Our solution enables high-quality load shedding in a multi-operator application. It therefore uses the whole application's stream properties, not only those at the bottleneck operator.

To the best of our knowledge, this is the first solution to have this perspective for CEP applications. Hence, to evaluate our solution's quality and to understand how much a global perspective can enhance the load-shedding quality, we compare our global solution with a \textit{local} solution.

Both solutions use the proposed LP.

The local strategy maximizes the sum of $\omega_2$'s output events. The global strategy maximizes the sum of events arriving at the sinks.

\subsection{Set Up}
For all evaluations, we use the implementation of the running example (Figure \ref{fig:google_borg_application} . The operator graph has four operators. Operators 1 and 2 implement two patterns each, Operators 3 and 4 one pattern. The graph further has two sources and two sinks.

\paragraph{Data Set}
To understand how the global shedder improves quality over the local shedder, we run implementation with different datasets.

A synthetic balanced data set where $\omega_1$ and $\omega_2$ emit a similar output rate per type. Here, we expected both local and global to generate similar quality. The lack of global knowledge is unimportant if $\omega_1$ emits the same number of outputs per type as $\omega_2$.

Further a real-world data-set from google borg\footnote{\url{https://github.com/google/cluster-data}, visited September 2023}. We map each measurement's machine-id to a number between 0 and 3. We have no control over the values' distributions or variances. In this real-world data case, we expected a higher quality from global shedding. 
The data set consists of multiple individual files. Hence, for each experiment, we feed a share of the files to Sink 1, another share to Sink 2. 

\paragraph{Average processing time as latency bound}
As explained in Section \ref{sec:solution}, we control an operator's latency by controlling the operator's average processing time $p_\omega$. Shedding reduces this average. Shed events have a processing time of 0. A queue at the operator is thus processed faster, enabling faster processing of non-shed events.
In the experiments, we therefore use $p_\omega$ as threshold to be met by shedding. 
\paragraph{Parameters and Thresholds}
For all evaluations, we permit a small deviation from the required processing time to avoid update-flooding under small changes. We chose a 10\% deviation for the processing time and a 5\% deviation for update messages. Update messages are emitted when an operator's measured processing time or output rates deviate from the latest reported value.

 \subsection{Evaluation of Output Quality and Latency Compliance}
 We evaluate output quality and latency compliance for local and global shedding in the following.
 
 \paragraph{Output quality}
\begin{figure}
	\includegraphics[width = \linewidth]{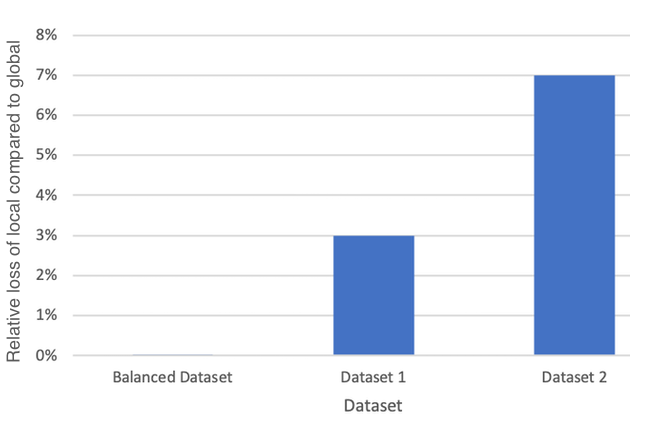}
	\caption{Relative loss in QoR of local compared to global shedding. \label{fig:quality}}
\end{figure}
Figure \ref{fig:quality} shows how many events are missed when using the local shedder in relation to shedding with the global shedder. We computed $1-(\sum_{Sinks}output_{local}/\sum_{Sinks}output_{global})$ to measure the relative loss of events by local compared to global.
For the balanced dataset, we see both shedders yield the same output, hence no loss is visible. For two datasets taken from the google borg files, we see that global shedding outperforms local in terms of quality. There is a relative loss in events in both cases.

\begin{figure}
	\includegraphics[width = \linewidth]{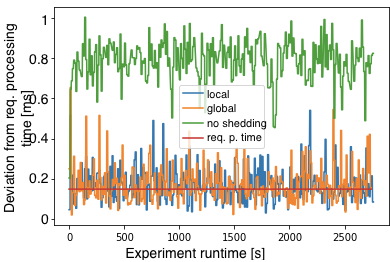}
	\caption{Latency compliance of the approaches\label{fig:latency_rt}}
\end{figure}

\begin{figure}
	\includegraphics[width = \linewidth]{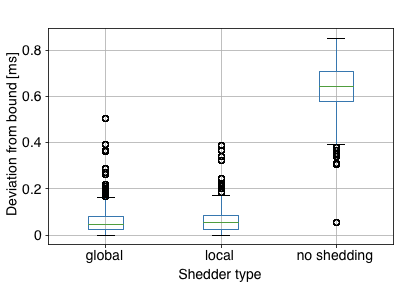}
	\caption{Latency compliance of the approaches\label{fig:latency}}
\end{figure}

 \paragraph{Latency compliance}
To understand whether both solutions comply with the required processing time, we further analyzed the processing times and present the results in two perspectives, displayed in Figures \ref{fig:latency_rt} and \ref{fig:latency}. 
Both figures shows that both shedding strategies are able to comply with the required processing times and reduce the average processing time compared to no shedding. 

Figure \ref{fig:latency_rt} shows how the processing time changes throughout runtime. Hence it shows that the solution is able to continuously adapt with low deviation.
Figure \ref{fig:latency} shows the total deviation in seconds of the solutions from the required processing time.  We measured to what extent the average processing time deviated from the required value. This includes both whether it exceeded or whether it was too low. 
The latter is counted if shedding is active. This captures the scenario where the shedder also recomputes a potentially non-optimal solution that sheds more than necessary. This situation can arise if stream characteristics change. 
We see that both shedding solutions - global and local - are able to keep deviations from the required processing time low, while the solution that did not shed has a higher value. The deviation is explained by the permitted threshold as explained before.

\subsection{Evaluation of the Solution's Scalability}
\begin{figure}
	\includegraphics[width = \linewidth]{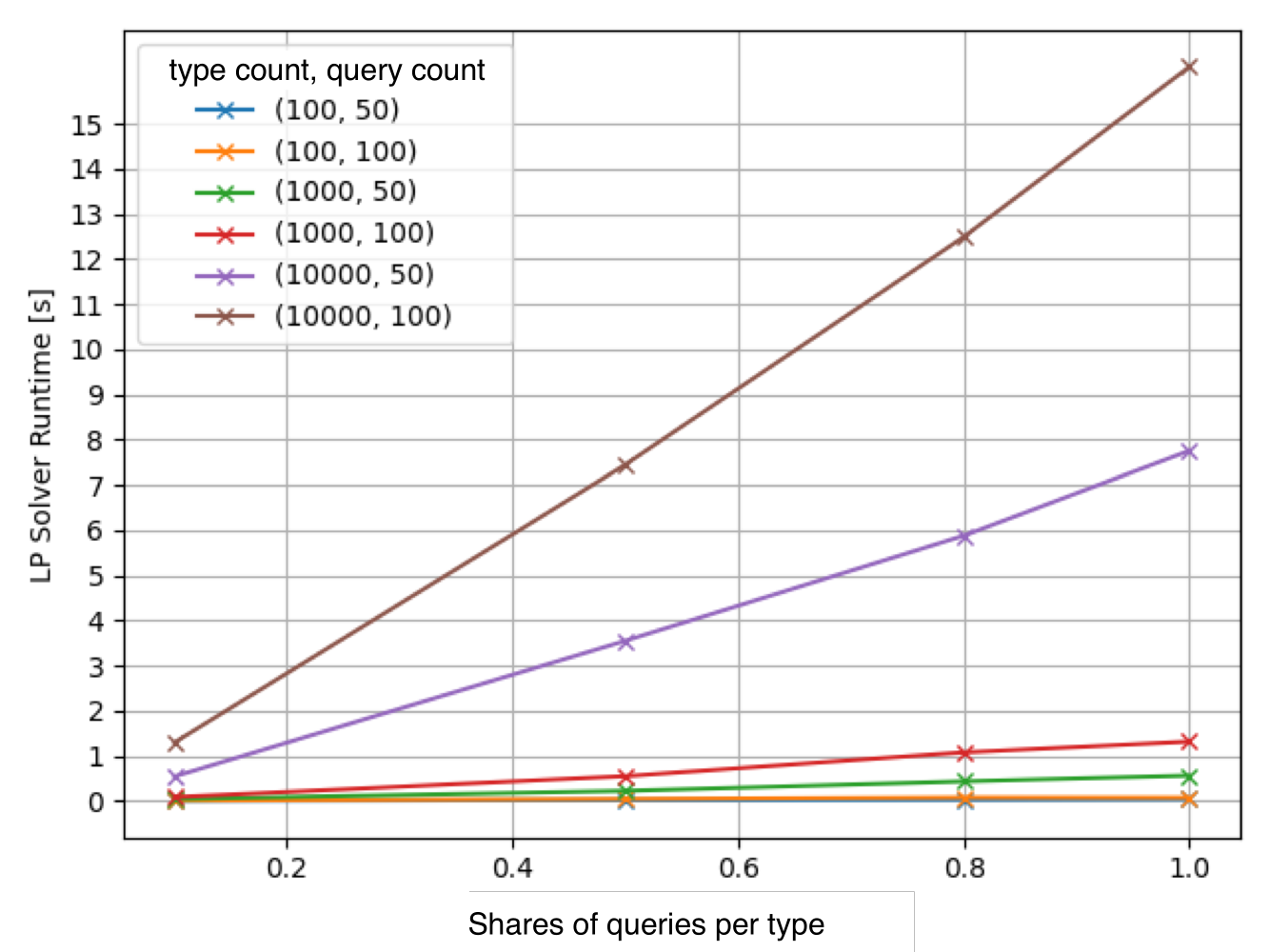}
	\caption{LP-solver runtime for different sizes of queries and type.\label{fig:scalability}}
\end{figure}

One objective is that in case of a new shedder configuration, the solver quickly finds a solution to prevent further latency violations. 
We, therefore, further evaluated the runtime of the LP-solver for different amounts of types, queries, and shares of queries per type. The later tells how many of the queries use the type.
Figure \ref{fig:scalability} shows that the latency-driving factor is the share of queries per type, i.e., in how many queries a type is present. 

We see that the solution is found below one second for all queries with up to 1000 types and 50 queries and only slightly exceed it for 1000 types and 100 queries. For 10.000 types, solver runtime grows up to 16.25 seconds at 100 queries, if all 10.000 types appear in all 100 queries.  

\subsection{Summary of Evaluations}
To sum up, we can state that both the local and the global LP-based shedding strategy are able to stick to a required latency bound. For a non-balanced, real-world dataset, the global strategy outperforms the local strategy in terms of quality. Finally, the solution proves to be highly scalable.

\section{Related Work}

While there are load shedding solutions available for single-node applications \cite{slo2019espice, slo_pspice_2019, katsipoulakis2018concept}, few are available for the distributed setting introduced. 
Solutions for CEP applications stem from Slo et al  \cite{slo2019espice, slo_pspice_2019} and Zhao\cite{zhao_load_2020}, yet are single-operator solutions, only. Load shedding for changing workloads is considered by Katsipoulakis et al.\cite{katsipoulakis2018concept}, yet again for single-operator and stream processing, only. 
Tatbul et al. \cite{tatbul_staying_2007} propose load shedding in distributed stream processing for multiple operators. 
Yet they do not consider conditional operators as given in distributed CEP.

Load shedding in stream processing, e.g., by Katsipoulakis et al.\cite{katsipoulakis2018concept} meet dynamics in the input-stream distribution with frequent sampling from arriving windows while considering a single operator. They support the claim that in stream processing, dynamic selectivities can often be found, for example for join operators and aggregations that can be utilized.
Tatbul et al. \cite{tatbul_staying_2007}  look at single- or multiple operators and pre-process, using selectivity functions. They model the problem as a linear optimization problem and offer both a solver-based centralized approach and a distributed approach. Both solutions rely on generating a series of load shedding plans in advance,to be used under certain load conditions. Again, the stream processing solutions do not have to consider the complexities of CEP operators.

Slo et al. \cite{slo2019espice, slo_pspice_2019} focus on load shedding with a focus on pattern detection.  
They learn an events utility based on its position, type, or the operator's internal state. The latter requires knowledge of this internal state, e.g., of the available partial matches at the operator. Further, they use the position of an event in a window as important property. However, in a multi-operator scenario, a shedder cannot tell at which position output events of its operators can end up for downstream operators.
For single and even complex operators, these are useful solutions. However, in a distributed setting, the used utility values might fall short as introduced before. 

In approximate processing  only a share of the input stream is processed. While the perspective focuses on what to process, not on what to drop, the outcome is similar. However, approximate processing can be found for stream processing, only \cite{FarhatBD20, GEDIK2016106}.

\section{Summary and Outlook}
In distributed CEP application, single operators can become a bottleneck which slows down the end-to-end latency of the whole application. In particular with limited resources, load shedding is a proven strategy to ensure timely processing of important events while shedding unimportant ones. In multi-operator CEP applications, a load shedder that maximizes the application's total output faces needs to consider not just its local input but the current input- and processing at all operators in the application. 
With the solution proposed in this paper, we enable load shedding in geographically distributed CEP applications. We thereby, in particular, introduce conditional selectivity functions for common building blocks of CEP queries. Using those selectivity functions, we further propose an LP formulation to maximize the application's output, given the current stream characteristics of the application. We finally showed in evaluations that our solution outperforms a local solution on real data sets, while maintaining latency bounds. Moreover, our solution proves to be highly scalable 
As future work, we look into more fine-grained type definitions and load shedding under probabilistic attributes. 

\bibliography{references}
\bibliographystyle{plain}

\end{document}